\documentclass[aps,twocolumn,prl,showpacs,amsfonts,amssymb,amsmath]{revtex4-1}
\usepackage{verbatim}
\usepackage{enumerate}
\DeclareMathAlphabet{\mathpzc}{OT1}{pzc}{m}{it}
\usepackage{braket}
\usepackage{graphicx}
\usepackage[dvipsnames]{xcolor}
\usepackage{import}
\usepackage{amsthm}
\usepackage{hyperref}
\usepackage{bm}
\usepackage{color}
\usepackage[normalem]{ulem}
\usepackage{epstopdf}

\usepackage{float}
\usepackage[caption = false]{subfig}

\hypersetup{colorlinks=true}
\hypersetup{citecolor=blue}
\hypersetup{linkcolor=blue}
\hypersetup{urlcolor=blue}

\providecommand{\abs}[1]{\lvert#1\rvert}

\newcommand{\E}{\mathrm{e}}
\newcommand{\I}{\mathrm{i}}
\newcommand{\D}{\mathrm{d}}

\begin{document}
	
	\title{Huygens-Fresnel principle: Analyzing consistency at the photon level}
	\author{Elkin A. Santos}\thanks{elkin.santos.s@gmail.com}
	\author{Ferney Castro}
	\author{Rafael Torres}
	\affiliation{Grupo de \'optica y tratamiento de se\~nales, Escuela de F\'isica, Facultad de Ciencias, Universidad Industrial de Santander, Bucaramanga, Colombia 680002}
	\date{\today}
	
	\begin{abstract}
	
Typically the use of the Rayleigh-Sommerfeld diffraction formula as a photon propagator is widely accepted due to the abundant experimental evidence that suggests that it works. However, a direct link between the propagation of the electromagnetic field in classical optics and the propagation of photons where the square of the probability amplitude describes the transverse probability of the photon detection is still an issue to be clarified. We develop a mathematical formulation for the photon propagation using the formalism of electromagnetic field quantization and the path-integral method, whose main feature is its similarity with a fractional Fourier transform (FrFT). Here we show that, because of the close relation existing between the FrFT and the Fresnel diffraction integral, this propagator can be written as a Fresnel diffraction, which brings forward a discussion of the fundamental character of it at the photon level compared to the Huygens-Fresnel principle. Finally, we carry out an experiment of photon counting by a rectangular slit supporting the result that the diffraction phenomenon in the Fresnel approximation behaves as the actual classical limit.
\end{abstract}

\maketitle	
	

	\section{Introduction}
	
In the scalar diffraction theory, the propagation of the electromagnetic field is formulated as a solution of the Rayleigh-Sommerfeld diffraction formula\cite{sommerfeld,bornwolf}, representing the Huygens-Fresnel principle, which can be simplified into the Fresnel diffraction approximation.
	
Generally, in quantum mechanics and quantum field theory, a $K(xt;x't')$ propagator is a Green's function representing the probability amplitude for a system to be in the position and time $(x',t')$ and at a later time, to be in a position and time $(x,t)$. The evolution of the system is expressed as
	\begin{equation}
	\Psi(x,t)=\int K(xt,x't')\Psi(x',t')\,\D^3x'\,, \label{eq:propagador}
	\end{equation}	
and these propagators can be studied within the framework of Feynman's path-integral formulation of the nonrelativistic quantum mechanics\cite{feynman,feynman-hibbs}. The concept of path integrals can be extended heuristically to the case of quantum electrodynamics, where the Feynman's propagator is now interpreted as the transition amplitude that a particle is created and destroyed by interaction. 
	
The interest lies in finding a propagator suitable for quantum optics, which allows us to formulate the propagation of a photon from one point to another. In his theory of photodetection\cite{glauber}, Glauber defines the detection of a photon by an absorption process, where the function $ \Psi(\vec{r},t) $ associated to the state $ \ket{\Psi} $ may be regarded as an ``electric-field wavefunction,'' sometimes called {\em effective wavefunction}, representing the probability amplitude of having a photodetection event at space-time point $ (\vec{r}, t) $. 	
	
Notwithstanding, experimental evidence that shows single-photon interference patterns\cite{kocsis} such as in classical optics, a clear quantum-formalism where the quantum propagator, for a large number of quanta, leads to the classical electromagnetic formalism is yet to be exposed and, according to experiments, the propagation of photons is dealt with through the scalar diffraction theory of classical optics\cite{saleh00,thiel09}. Furthermore, since a proper wave function for photons is still a highly arguable topic\cite{sipe,birula1,hawton3,smithraymer}, there is no way to use the Born's interpretation of the wave function\cite{born} to directly relate  the probability density of detection with the diffraction pattern obtained when the electromagnetic field is propagated classically.

Using a non-relativistic approach, our development is based on each mode of the radiation field being treated as independent quantum oscillators, $\hat{H}=\frac{1}{2}(\hat{p}^2+\omega^2\hat{q}^2)$ \cite{dirac,BJ,BJH}. By calculating the transition amplitude associated with such Hamiltonian, the Feynman propagator of the canonical position $q$ is constructed. Here, the observable $ q $ will be thought of, using the proper scale factors, as a position coordinate  perpendicular to the direction of propagation of the field. Also, given the close relation between the fractional Fourier transform (FrFT) with harmonic systems\cite{namias,kutay}, and especially with the propagation of the electromagnetic field\cite{pellat,pellat2,cai05,tasca09}, we show that the propagation of photons takes the form of the classic Fresnel diffraction integral instead of the Huygens-Fresnel principle, in which the spherical wave fronts are replaced by paraboloidal wave fronts that cannot be generated by point sources.

Some experiments in classical optics (see Refs.\cite{sheppard,southwell,steane}) have shown that the Fresnel approximation is surprisingly accurate, even in regions very close to the diffraction aperture where the corresponding approximation should no longer be valid (see discussion in \cite{goodman}). Then, using an approach where sources have finite dimension instead of point sources, we show that the Helmholtz-Kirchhoff equation can also be solved satisfying either the Dirichlet or Neumann boundary conditions, and using a suitable distribution function the Fresnel diffraction integral can be obtained directly.

	\section{Photon propagator}
		
Consider the Hamiltonian of the quantized electromagnetic field, $\hat{H}=\frac{1}{2}(\hat{p}^2+\omega^2\hat{q}^2)$, which is formally equivalent to a mechanic harmonic oscillator, where $ q(t) $ and $ p(t) $ play the roles of canonical position and momentum, having dimensions of $ [m\sqrt{Kg}] $ and $ \left[\frac{m}{s}\sqrt{Kg}\right] $, respectively. The corresponding transition amplitude for this Hamiltonian computed via the path integral method\cite{feynman-hibbs} is
\begin{align}
\bra{q_F}\E^{-\I\hat{H}/\hbar}&\ket{q_I}=\left( \frac{\omega }{2 \pi\I\hbar \sin \omega t}\right) ^\frac{1}{2}\\
&\times\exp\left[\frac{\I}{2\hbar}\omega\left(\left[q_I^2+q_F^2\right]\cot\omega t-\frac{2q_Iq_F}{\sin\omega t}\right)\right]\,.\nonumber
\end{align}
Then, the temporal evolution of a system in the state $|\psi\rangle$ in the $ q $ representation is written
\begin{align}\label{eq:propfoton1}
\psi(q_F,t)=&\left( \frac{\omega }{2 \pi\I\hbar \sin \omega t}\right) ^\frac{1}{2}\int_\mathbb{R} \,\psi(q_I,0)\\
&\times\exp \left[ \frac{\I}{2 \hbar} \omega \left( \left[ q_I^2 + q_F^2 \right] \cot \omega t \right.\right.
\left.\left.- \frac{2q_Iq_F}{\sin \omega t} \right) \right]\,\D q_I\,.\nonumber
\end{align}
This equation describes how the ``wave function'' evolves in time as light propagates, whose kernel is written as
 a Fourier transform and a quadratic phase just like the Fresnel diffraction integral in the classic electromagnetic field propagation\cite{sommerfeld,goodman,bornwolf}.  This similarity allows us to establish a connection where the Fresnel diffraction integral plays a major role in the photon propagation.

\section{Clasical electromagnetic field propagator}

\begin{figure}[h!]
\includegraphics[scale=1]{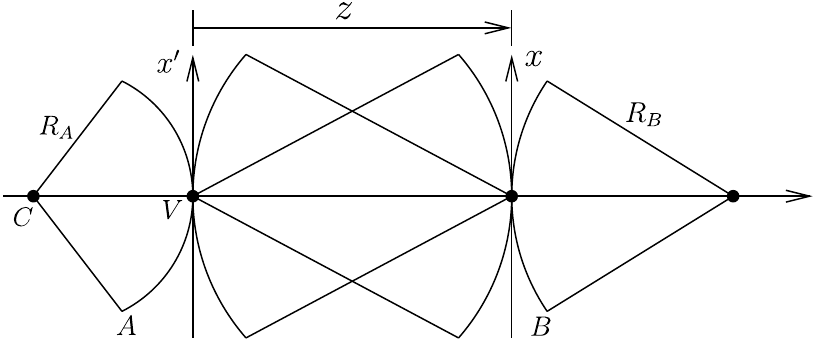}
\caption{Fresnel diffraction between the spherical surface $A$ to spherical surface $B$.}\label{figmet}
\end{figure}

The one-dimensional (1D) Fresnel diffraction integral to distance $z$, in the framework of the Bonnet metaxial optics, is written
\begin{align}\label{key3}
&U(x,z)=\left(\frac{\I}{\lambda z}\right)^{\frac{1}{2}}e^{\I kz}\exp\left[{-\frac{\I k}{2}\left(\frac{1}{z}+\frac{1}{R_{B}}\right)x^2}\right]\\
&\times\int_\Sigma \exp\left[-{\frac{\I k}{2}\left(\frac{1}{z}-\frac{1}{R_{A}}\right)x'^2}\right]\exp\left[{\frac{\I k}{z}xx'}\right]U(x',0)\D x'\,,\nonumber
\end{align}
where the spherical waves are approximated into parabolic ones and the radius of the curvature for $U$ (see Fig.\ref{figmet}) from its vertex to the center is defined as the algebraic quantity $R_{A}=\overline{VC}$, and it is considered positive if it goes in the direction of propagation of light. 

Pellat-Finet\cite{pellat,pellat2} established a relationship between the Fresnel diffraction and the fractional Fourier transformation\cite{namias}.  So, defining
\begin{equation}\label{key1}
		\cot\alpha=\varepsilon\dfrac{1-\mu}{\mu}\,,\quad\quad\text{with}\quad\quad \mu=\dfrac{z}{R_A}\,,
		\end{equation}
then	
		\begin{equation}
		\sin^{2}\alpha=\frac{\mu^{2}}{\mu^{2}+\varepsilon^{2}(1-\mu)^{2}}\,,\label{eqss}
		\end{equation}
where $\varepsilon$ is the real-number nonzero solution of
		\begin{equation}
		\frac{1}{R_{B}}+\frac{1}{z}=\frac{\varepsilon^{2}(1-\mu)}{\mu R_{A}[\mu^{2}+\varepsilon^{2}(1-\mu)^{2}]}\,.\label{eqR}
		\end{equation}
Thus, with the following choice of reduced variables:
		\begin{equation}\label{key2}
		\rho=\dfrac{1}{\sqrt{\lambda\varepsilon R_A}}x'\,,\quad \sigma=\dfrac{1}{\sqrt{\lambda\varepsilon R_A}}(\cos\alpha+\varepsilon\sin\alpha)x\,,
		\end{equation}
and reduced amplitudes,
\begin{equation}
V_{A}(\rho)=U_{A}(\sqrt{\lambda\varepsilon R_A}\rho)\,,\quad V_{B}(\sigma)=U_{B}\left(\dfrac{\sqrt{\lambda\varepsilon R_A}\sigma}{\cos\alpha+\varepsilon\sin\alpha}\right)\,,
\end{equation}
Eq.\ref{key3} is written in the form
\begin{align}
V_{B}(\sigma)&=\frac{\I}{\sin\alpha}(\cos\alpha+\varepsilon\sin\alpha) \nonumber \\
\times&\int_{\mathbb R}V_{A}(\rho)\exp{\left(\I\pi[\rho^{2}+\sigma^2]\cot\alpha-\frac{2\I\pi\sigma\rho}{\sin\alpha}\right)}\D\rho\,.\label{eqfresnel}
\end{align}
Now the relationship between the photon propagator \ref{eq:propfoton1} and the Fresnel diffraction integral written in the form of Eq.\ref{eqfresnel} is clearer. We just need to find the scale factor for the appropriate reduced variables for $q_{I}$ and $q_{F}$.
	
\section{Propagation as a Fresnel integral}
We establish a relation between our propagator \ref{eq:propfoton1} and the fractional Fourier transform\cite{namias,mcbride} by taken the following change of variables:
\begin{equation}
\rho^{2}=\frac{\omega}{2\pi\hbar}q_I^{2}\,,\qquad 	\sigma^{2}=\frac{\omega}{2\pi\hbar}q_F^{2}\,.
\end{equation}
Using $\alpha=\omega t$, $k=\frac{2\pi}{\lambda}$ and $ct=\varepsilon R_{A}>0$, we arrive at
\begin{equation}
\rho^{2}=\frac{c\alpha}{\lambda\varepsilon R_{A}\hbar k}q_I^{2}\,,\qquad \sigma^{2}=\frac{c\alpha}{\lambda\varepsilon R_{A}\hbar k}q_F^{2}\,.
\end{equation}\\
We define the ``mass'' term $m_{\lambda}=\frac{\hbar k}{c}$, which is related to the Hamiltonian, that is, the quantized electromagnetic field can be understood as a quantum mechanical harmonic oscillator with ``mass'' $m_{\lambda}=\frac{\hbar k}{c}$, so
	\begin{equation}
	\rho^{2}=\frac{1}{\lambda\varepsilon R_{A}}\frac{\alpha}{m_{\lambda}}q_I^{2}\,,\qquad 	\sigma^{2}=\frac{1}{\lambda\varepsilon R_{A}}\frac{\alpha}{m_{\lambda}}q_F^{2}\,.
	\end{equation}
Then, by using the expressions in \ref{key2}, we find the scaling between the observable $ q $ and a real position $ x $ to be 
\begin{equation}
x_{I}^{2}=\frac{\alpha}{m_{\lambda}}q_{I}^{2}\,, \quad x_{F}^{2}=\frac{\alpha}{m_{\lambda}}q_{F}^{2}\,,
\end{equation}
where 
\begin{equation}\label{eq:epsilon}
	\cos\alpha+\varepsilon\sin\alpha=1.
\end{equation}

Then we can write \ref{eq:propfoton1} in the form
\begin{align}
	\Psi(x_F,&t)=\left( \frac{\omega }{2 \pi\I\hbar \sin \omega t}\right) ^\frac{1}{2}\left (\frac{m_\lambda}{\alpha}\right )^\frac{1}{2}\int_\mathbb{R} \,\Psi(x_I,0)\label{eq20}\\
	&\times\exp \left[ \frac{\I\pi}{\lambda\varepsilon R_{A}}\left( \left[ x_I^2 + x_F^2 \right] \cot \alpha - \frac{2x_Ix_F}{\sin \alpha} \right) \right]\,\D x_I\,,\nonumber
	\end{align}
where $ \Psi(x_I,0)=\psi\left (\sqrt{\frac{m_\lambda}{\alpha}}x_I,0\right ) $ and  $ \Psi(x_F,t)=\psi\left (\sqrt{\frac{m_\lambda}{\alpha}}x_F,t\right ) $.\\

In addition, by using \ref{key1} and \ref{eq:epsilon} we arrive at
\begin{equation}
\sin\alpha=\frac{\mu}{\varepsilon}\,,\label{eqs}
\end{equation}
So, equating \ref{eqs} and \ref{eqss} we have
\begin{equation}
\mu^{2}+\varepsilon^{2}(1-\mu)^{2}=\varepsilon^{2}\,,
\end{equation}
with $\varepsilon^2=\frac{\mu}{2-\mu}$. Then \ref{eqR} takes the form
\begin{equation}
\frac{1}{R_{B}}+\frac{1}{z}=\frac{1-\mu}{\mu R_{A}}\,.
\end{equation}
Finally, using \ref{key1}, we have $R_{B}=-R_{A}$ and the expression \ref{eq20} can then be written explicitly in terms of the position and the propagation distance $ z $ as
\begin{widetext}
	\begin{equation}\label{eq:propdensi}
		\Psi(x_F,z)=\left(\frac{\I}{\lambda z}\right)^{\frac{1}{2}}\exp\left[-{\frac{\I k}{2}\left(\frac{1}{z}+\frac{1}{R_B}\right)x^2}\right]
		\times\int_\Sigma \exp\left[{-\frac{\I k}{2}\left(\frac{1}{z}-\frac{1}{R_A}\right)x_F^2}\right]\exp\left[{\frac{\I k}{z}x_Ix_F}\right]\Psi(x_I,0)\D x_I\,,
	\end{equation}	
\end{widetext}
which is exactly the classical Fresnel diffraction formula dropping the phase factor $ e^{ikz} $.

Also, it can be written in terms of the reduced variables $ \rho $ and $ \sigma $ as the fractional Fourier transform,

	\begin{align}\label{eq:FrFT}
	\phi(\sigma)=&\left( \frac{1}{\I\sin\alpha}\right) ^\frac{1}{2}\int_\mathbb{R} \exp \left[\I\pi\left( \left[\rho^2 +\sigma^2\right] \cot\alpha - \frac{2\rho\sigma}{\sin \alpha} \right) \right]\nonumber\\
	&\times\phi(\rho)\,\D\rho\,.
	\end{align}

Thus, the FrFT mathematically expresses the photon propagation in the same way it is used to propagate the classical field in the Fresnel regime. Note that when $\alpha=\pi/2$, the propagation becomes the standard Fourier transform (Fraunhofer regime). What is remarkable here is that, now we have a well known tool to study the propagation of photons, and we can apply all the properties of Fourier analysis to quantum optics.

As we can see, the wave function $ \Psi(x,t) $ behaves as an electric-field wave function that is closely related by some scale factors to the wavefunction in the $ q $-representation $ \psi(q,t) $ and can be propagated in the same way as the classical Fresnel diffraction integral. Since in both cases--quantum and classic--the probability amplitude and the electric-field amplitude evolve in the same way, this means that the observable $ q $ can be considered as a position in the transverse direction to the field propagation and parallel to the direction of electric-field polarization. 

Thus, Eq.\ref{eq:propfoton1} would represent the evolution of the transverse probability amplitude, $ \psi(q,t) $, of detecting a photon, remaining delocalized longitudinally. This means that the $ \hat{q} $ observable is far from what can be regarded as the position of the photon\cite{hawton1,hawton2,birulas}. It is worth mentioning that there is no position operator for photons \cite{pauli,newton} and there is not a satisfactory quantum-mechanical description for the photon in the usual sense.

		\section{Correspondence to the scalar diffraction theory}
Is it possible that the Fresnel diffraction is not a mere approximation of the Rayleigh-Sommerfeld formula but describes the propagation of the radiation field in a fundamental way? Let us see how the scalar diffraction theory can be adjusted in order to obtain a different expression of a propagated field.

	We propose that the electromagnetic field cannot be confined into a point region but in a small volume. It is not very instinctive to think of a point source for an electromagnetic wave or photons since the spatial energy density would be infinite. For instance, in Ref.\cite{romero}, the authors demonstrate that Huygen's secondary-sources have finite dimension and energy density; also, it has been shown\cite{birulas} that photons cannot be sharply localized,  although the possibility of having zero-area single photon pulses has been studied\cite{costanz}. Therefore, we consider that any source of electromagnetic waves $ U(\vec{r},t) $ with wavelength $\lambda$ must have a constant amplitude in a neighborhood of at least the order of the wavelength.

\begin{figure}[h]
	\subfloat[]{\includegraphics[width=8.6 cm]{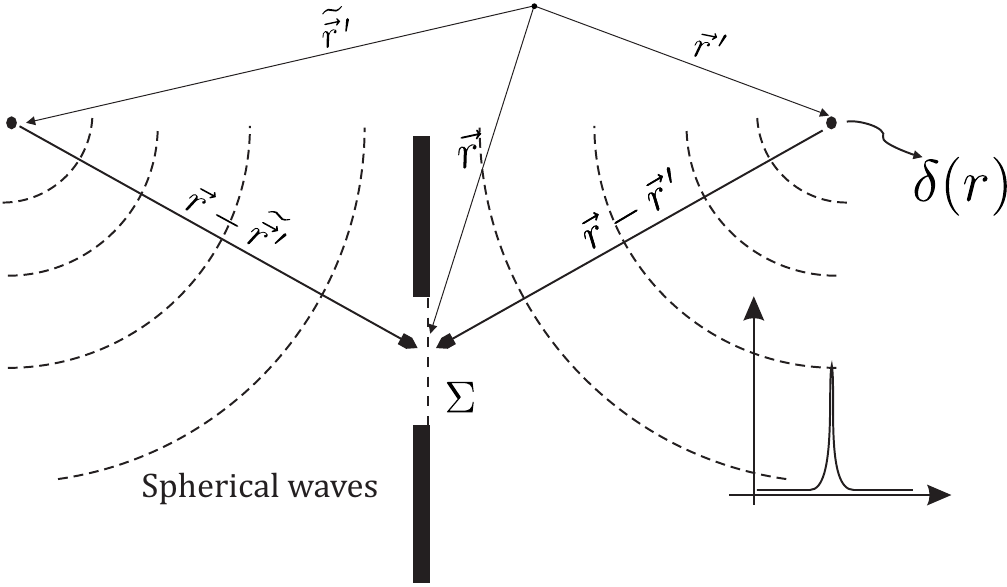}}\\
	\subfloat[]{\includegraphics[width=8.6 cm]{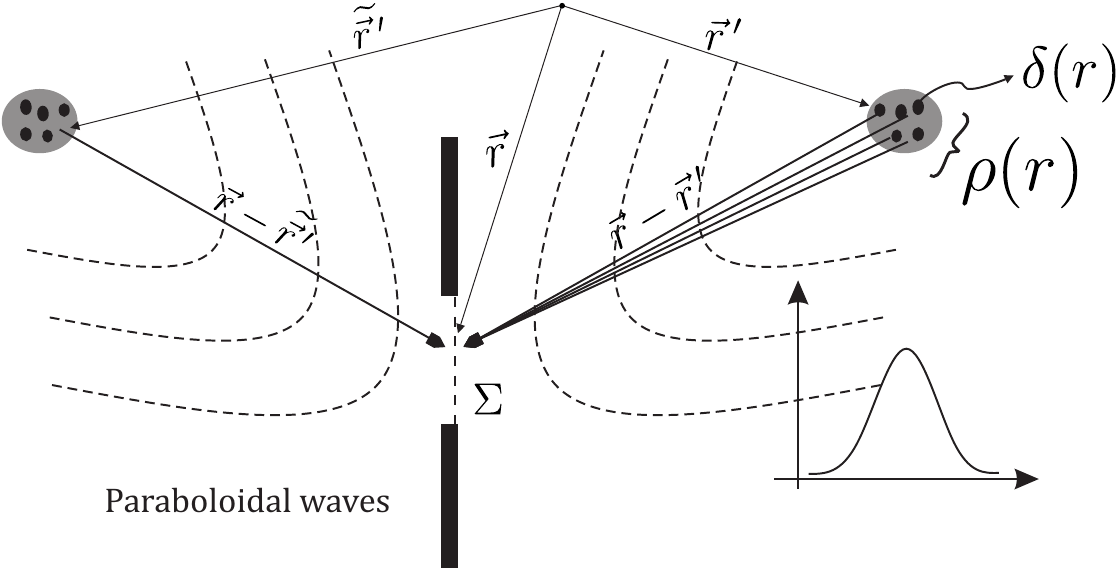}}
	\caption{Green's problem using a Dirac $\delta$ distribution. Proposed problem for an arbitrary distribution.}
	\label{fig:GyS}
\end{figure}

	Now, let us recall that in the scalar theory of diffraction, the Green's theorem is used to calculate the propagation of the electromagnetic field \cite{sommerfeld,goodman,bornwolf}. It is desired to solve, for $U$, the expression
	\begin{equation}\label{eq:Th Green}
	\int_VU(\nabla^2+k^2)G\D v=\int_S\left(U\frac{\partial G}{\partial n}-G\frac{\partial U}{\partial n}\right)\D s\,,
	\end{equation}
	where $G$ is an auxiliary function, called the Green's function,
	\begin{equation}
	(\nabla^2+k^2)'G_{\pm}(\vec r, \vec r\,')=-4\pi[\delta(\vec r-\vec r\,')\pm\delta(\vec r-\widetilde{\vec r\,'})]\,,\label{eqG}
	\end{equation}
	which represents a point source in $\vec r\,'\in V$ and $\widetilde{\vec r\,'}\notin V$. This allows the calculation of the field $ U $ in the region $V$ of space at the right of the plane $\Sigma$ [Fig.\ref{fig:GyS}, (a)]. One solution for $G$, in the sense of distributions, corresponds to the spherical wave,
	\begin{equation}\label{EqS}
	G_{\pm}(\vec r,\vec r\,')=\frac{e^{\I k\abs{\vec r-\vec r\,'}}}{\abs{\vec r-\vec r\,'}}\pm\frac{e^{\I k\abs{\vec r-\widetilde{\vec r\,'}}}}{\abs{\vec r-\widetilde{\vec r\,'}}}\,.
	\end{equation}
	Accordingly, the expression \ref{eq:Th Green} is reduced to
	\begin{equation}
	U(r)=\frac{1}{4\pi}\int_\Sigma U\frac{\partial G_{-}}{\partial n}\D s\,,
	\end{equation}
	which gives rise to the Rayleigh-Sommerfeld formula of the Huygens-Fresnel principle. 
	
	Now, since our premise is that an electromagnetic wave cannot be defined at a single point, but distributed in a neighborhood $ v(\vec r)\subseteq V$ of a point defined by $\vec r$, the Dirac distribution should be replaced by another distribution that allows us to consider the field in the neighborhood $v(\vec r)$ as a constant  $ U(v(\vec r))=U(\vec r) $ [Fig.\ref{fig:GyS}, (b)].
	
	Thus, it is proposed, instead of Green's condition (Eq. \ref{eqG}), a new condition
	\begin{equation}
	(\nabla^2+k^2)'\mathcal{G}_{\pm}(\vec r, \vec r\,')=\rho(\vec r-\vec r\,')\pm\rho(\vec r-\widetilde{\vec r\,'})\,,\label{eqR2}
	\end{equation}
	where the expression
	\begin{equation}\label{eq:cond}
	\int_VU(\vec r\,')\underbrace{(\nabla^2+k^2)'\mathcal{G}_{\pm}}_{\rho(\vec r-\vec r\,')}\,\D v'=U(v(\vec r))=U(\vec r)
	\end{equation}
	still holds. Recall that the integral is computed in the region $V$ [right hand side of scheme (\ref{fig:GyS})], where the auxiliary mirror source does not affect the result.

	Let us note that if we place point sources in pairs (one in the region where the field is measured and another in the mirror image), one pair after the other until a cluster is formed [Fig.\ref{fig:GyS}, (b)], then $\rho$ can be written in a basis of Dirac's distributions, and either the function $\mathcal{G}_{\pm}$ or its derivative can be chosen to be zero in the plane $\Sigma$ fulfilling either the conditions of Dirichlet or Neumann \cite{sommerfeld,goodman,bornwolf}. That is, $ \mathcal{G}(\Sigma)=0 $ or $ \frac{d\mathcal{G}}{dn}(\Sigma)=0 $ can be chosen, and then the field can be solved for a volume neighborhood $v(\vec r)$ (of the order of the wavelength).

	The function $ \mathcal{G}_{\pm}(\vec{r},\vec{r}') $, which is solution of Eq.\ref{eqR2}, is no longer a spherical wave and can be written in the form
	\begin{equation}\label{eq:sphe}
	\mathcal G(\vec r,\vec r\,')=\int_{v}\rho(\vec r\,''-\vec r)G_{\pm}(\vec r\,'',\,\vec r\,')\D\vec r\,''\,.
	\end{equation}
	Dropping the constant $ -4\pi $, we have
	\begin{align}\label{eqDem}
	(\nabla^2+k^2)'\mathcal G(\vec r,\vec r\,')&=\int_{v}\rho(\vec r\,''-\vec r)\\
	&\times(\nabla^2+k^2)'G_{\pm}(\vec r\,'',\,\vec r\,')\D\vec r\,''\,,
	\end{align}
	Using Eq. \ref{eqG} we have
	\begin{align}
	(\nabla^2+k^2)'\mathcal G(\vec r,\vec r\,')&=\int_{v}\rho(\vec r\,''-\vec r)\\
	&\times\left(\delta(\vec r''-\vec r\,')\pm\delta(\vec r''-\widetilde{\vec r\,'})\right)\D\vec r\,''\,,
	\end{align}
	which means that $ \mathcal{G} $ is composed of point sources along the neighborhood $v$, that is, a volume  of Dirac distributions.
	
	Then the solution for the field $ U $ is given by
	\begin{equation}
	U(r)=\frac{1}{4\pi}\int_\Sigma U\frac{\partial \mathcal{G}_{-}}{\partial n}\D s\,,
	\end{equation}
	and $ \mathcal{G}_{-} $ may be chosen to obtain any other solution for the diffraction integral provided that \ref{eq:cond} is valid.

Since the propagation for photons that we obtained is essentially the Fresnel diffraction integral formula where the waves are not spherical but paraboloidal, they cannot be associated with a Dirac $\delta$ distribution, but with a different type of distribution as shown before\footnote{Finding the distribution which leads directly to the Fresnel diffraction is beyond the scope of the paper, although the Gaussian distribution seems to be right option\cite{Wunsche}.}. That is, the paraboloidal wave fronts cannot be produced by point sources, but by sources with some dimension. Huygens' principle is, in this sense, a particular case of punctual sources that works fine when, in the neighborhood $ v $, the electromagnetic field can be approximated in classical theory by a point source diffracting light in all directions. In consequence, we could think of the solution of Eq.\ref{eqR2} in such a way that the new distribution leads to a new auxiliary function, $ \mathcal{G_\pm} $, where the Fresnel diffraction is obtained directly.

It has been shown\cite{Feiock,Grella} that the Fresnel diffraction is an exact solution of the paraxial wave equation and that the paraxial equation is also equivalent to the time-dependent Schr\"odinger equation\cite{nienhuis93} for a particle moving in a two-dimensional potential, where the $ z $ coordinate plays the role of time, and also has been used before to study the transverse localization of light\cite{Vries}. Our treatment would then justify the use of the Fresnel diffraction as a propagator for light quanta since it suggests that the propagator can be written as so based on a fully quantum approach.
	
\section{Photon counting experiment}
	
\begin{figure}[h!]
	\includegraphics[scale=1]{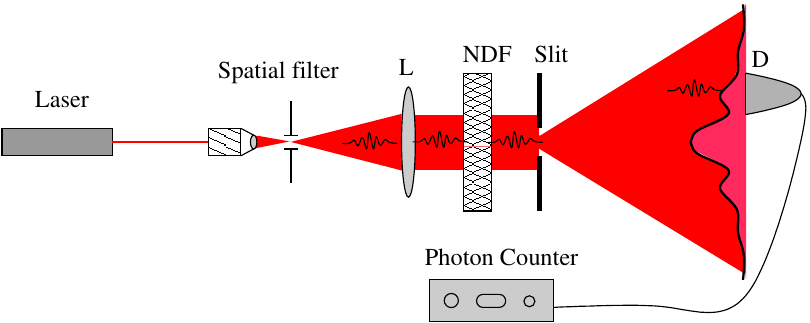}
	\caption{Experimental setup}\label{figexp}
	\end{figure}

The results found here are verified through the implementation of a diffraction experiment by photon counting (see figure \ref{figexp}). 
This was developed with the only intention of showing the correspondence between this photon propagator and the experiments, which justifies the use of it for example to study the correlation between entangled photons\cite{cai05,tasca09}.
\begin{figure}[h!]
\includegraphics{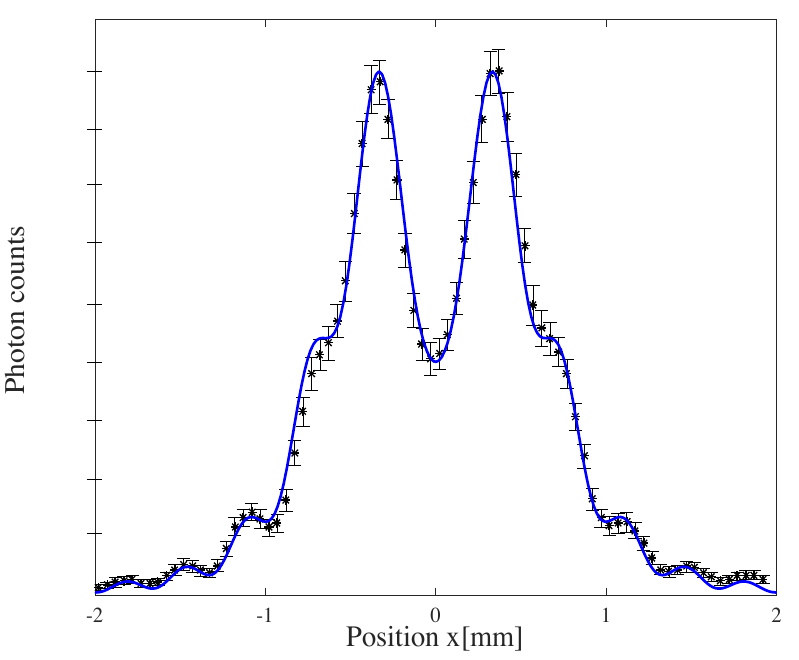}
\caption{The experimental data overlapped to the blue curve which is the normalized probability density, $ \abs{\Psi(x,z)}^2 $, obtained by computational simulation.}\label{figres}
\end{figure}
	
Our results are verified in a diffraction experiment with a laser beam ($\lambda = 632\,nm$) collimated with linear polarization, which is attenuated by means of an array of neutral density filter (NDF) until counting a limited number of photons and then the beam is diffracted by a rectangular slit of $1905\,\mu m$. The diffraction is made in propagation in the free space, at a distance of 96.84 cm from the slit to a photon counter of the avalanche photodiode (D), whose diameter is $50\, \mu m$. The data is taken during a time of $10\, ms$, by scanning the diffracted field.

The correct fit between the experimental data and the photon propagator according to Eq.\ref{eq:propdensi} (see Fig.\ref{figres}), reveals that the $ q $-representation of a state of the field can be used as a guidance to study the photon propagation.

 	\section{Simulation of the probability distribution for single photon propagation}
 	
\begin{figure}[h]
 		\includegraphics[scale=0.43]{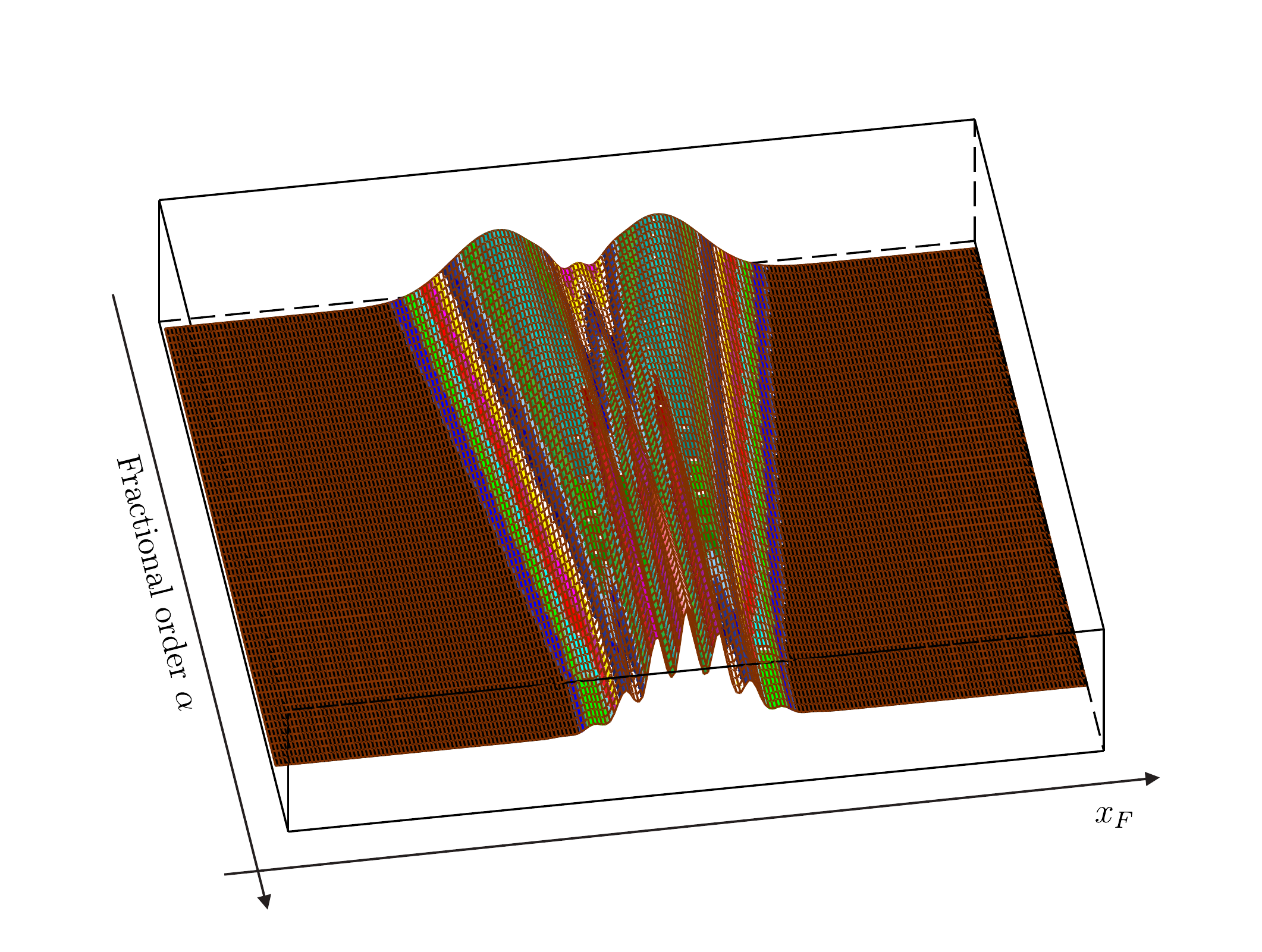}
 		\caption{Simulation of several probability density distributions at a distance, $ z $.}
 		\label{fig:kocsis}
 	\end{figure}

It is very interesting to explore how the probability density distribution evolves as the distance of the plane of observation increases in the double-slit experiment. Here we show a simulation for several planes of observation using the propagator in Eq.\ref{eq:FrFT}. As the order of the FrFT approaches to $ \alpha=\pi/2 $, the probability density distribution changes until it reaches the characteristic Fraunhofer diffraction pattern, as shown in Fig.\ref{fig:kocsis}. This is a Young interferometer using Gaussian beams as sources with waist $ 1/e^2 $ radius of $ 0.6 $ mm and peak-to-peak separation of $ 4 $ mm. The distribution densities are plotted from $ \alpha=0.8\pi/2 $ to $ \alpha=\pi/2 $ related to the propagation distance $ z $.
	
In our paper, the use of the Fresnel diffraction integral or, equivalently, the fractional Fourier transform is now fundamentally justified for single-photon propagation. Our treatment agrees with the probability density distributions experimentally found by Kocsis \textit{et al}.\cite{kocsis}, in which they were able to construct  classical trajectories for single photons in the double slit interferometer by means of a weak measurement of the momentum without destroying interference. That is, the overall conclusion, where those trajectories represent the average behavior of the ensemble of photons, is confirmed by our result.

 	\section{Summary}

In summary, there are several reasons that point in the direction that the Fresnel diffraction has a fundamental character and it is not only a mere approximation:\\
	(i) The propagator of the $ q $-representation of any state of the field can be written in the form of the classic Fresnel diffraction integral.\\	
	(ii) Spherical wave fronts can only be produced by point sources, contrary to the paraboloidal wave fronts that can only be produced by sources with some dimension.\\	
	(iii) Photons are not point particles since it would imply an infinity energy density.\\	
	(iv) Experiments show that the Fresnel diffraction integral is more accurate than expected, even in regions where it should no longer be valid.
	
These points suggest the following:

	\textit{Remark}. The wave front of a ``wave function'' for photons is composed for secondary sources with certain dimension which produce new parabolic waves that construct the new wave front, allowing the probability amplitude to propagate.

	\section{Conclusion}
	
We have found a propagator for photons that takes the form of the classical Fresnel diffraction integral, by means of the close connection between both of them and the fractional Fourier transform. We showed that the $ q $ observable, properly scaled, corresponds to a position observable transversal to the propagation of the field and parallel to the electric field polarization. 
	
This means that in the limit for large number of quanta, the classical intensity of the field is then proportional to the probability density, $ \abs{U(x,z)}^2\propto\abs{\psi(q,t)}^2 $, and so the correspondence principle is satisfied, as shown in our photon-counting experiment.
	
Finally, we also showed that the Green's problem in the scalar theory of diffraction can be adjusted by using a proper distribution that changes the spherical waves produced by the Dirac distribution into the characteristic paraboloidal waves of the Fresnel diffraction to obtain the latter, not as an approximation but as an exact result .

		\bibliographystyle{apsrev4-1}

\end{document}